\documentclass[a4paper,11pt]{article}
\pdfoutput=1 

\usepackage{jcappub} 

\usepackage[T1]{fontenc} 

\title{\boldmath Hyperon threshold and stellar radii}

\author[a,1]{Luiz Lopes,\note{Corresponding author.}}
\author[b]{Debora Menezes}

\affiliation[a]{ Centro Federal de Educa\c{c}\~ao Tecnol\'ogica de Minas Gerais Campus VIII; CEP 37.022-560, Varginha - MG - Brasil
}
\affiliation[b]{ Departamento de Fisica, CFM - Universidade Federal de Santa Catarina;  C.P. 476, CEP 88.040-900, Florian\'opolis, SC, Brasil 
}

\emailAdd{llopes@cefetmg.br}
\emailAdd{debora.p.m@ufsc.br}

\abstract{We study how the $\Lambda$ hyperon threshold influences the
  radius of the canonical $1.4 ~M_\odot$ neutron star in the light of
  the measurements found in the recent literature. We show that the onset
  of a new degree of freedom not only causes the well known reduction
  of the maximum mass, but also compacts the neutron stars with high
  central density. With the help of the strange mesons $\phi$ and
  $\sigma^*$, we show that it is possible to simulate very compact
  neutron stars keeping realistic hyperon potentials, $U_\Lambda(n_0)=
  -28$ MeV and  $ U_\Lambda^\Lambda(n_0/5)$ in agreement with recents measurements. 
   In the end we generalize these results showing that the onset
    of a yet not known dark matter particle with mass of 1.04 GeV is
    able to produce simultaneously a 2 $M_\odot$ neutron star and a
    canonical one with a radius of only 11.62 km. }

\arxivnumber{1701.03211}

\begin{document}
\maketitle
\flushbottom

\section{Introduction}
\label{sec:intro}

The physics of cold strongly interacting matter at extreme densities is
still not available in terrestrial laboratories. So far, the 
only place believed to reach such condition is the interior of neutron stars.
The observations of pulsars evolved in the last years and provided us more precise information about the macroscopic characteristics of these objects. 

For instance, the observations of two hyper massive pulsars, PSR J1614-2230~\cite{Demo} and PSR J0348+0432~\cite{Antoniadis},
show us that the EoS of beta-equilibrium nuclear matter should be very stiff, being able to produce a two solar masses neutron star.
This indicates that hadron interaction at very small distances  is strongly repulsive.

On other hand, the physics of neutron stars radii also improved in this decade. Recent observations indicate that the radii of
pulsars of mass around the canonical 1.4$M_\odot$ are lower than most
of the previsions found in the literature  \cite{Lattimer_2016}. Today we know 
that the radius of the neutron stars are somehow related to the symmetry energy slope, $L$~\cite{Rafa2011,Gandolfi,Rafa2014,Lopes14},
although $L$ cannot be the ultimate information about  neutron stars radii, once the same model with the same slope, predicts neutron 
stars  whose difference in radius reaches 1.7 km~\cite{Lopes14}. Theoretical works and astrophysical observations have strongly
constrained the radius of the canonical mass in the last couple of years. For example, based on a chiral effective theory,
ref.~\cite{Hebeler} constrained the radii of the  canonical
$1.4M_\odot$ neutron star to 9.7-13.9 km.
To fit experimental information from neutron skins, heavy ion
collisions, giant dipole resonances, and dipole polarizabilities,
ref.~\cite{Lim} constrained the neutron star radius of a canonical
mass in the narrow window 10.7 km $<$ R $<$ 13.1 km (90$\%$ confidence).
Using time-resolved spectroscopy of thermonuclear X-ray bursts observed from and object called SAX J1748.9-2021 ref.~\cite{Ozel}
constrained the radius as R = 10.93 $\pm$ 2.09 km for a mass M = 1.33 $\pm$ 0.33 $M_\odot$.From Fig. 4 of ref.~\cite{Steiner}
and Table 8 of ref.~\cite{Steiner2} a limit around 12.50 km is found for a
neutron star of mass 1.4$M_\odot$. Wee see that recent researches point to radii for the canonical 1.4$M_\odot$
 around 12 km. It should be however mentioned that current determinations of neutron star radius are subject to many uncertainties and systematic errors~\cite{Steiner3, Miller}

In this work we show that the onset of a new degree of freedom, not only causes the well known softening of the EoS but also
reduces the radii of the stars whose central density is higher than the density of the hyperon threshold (what we call here  subsequent stars),
 when compared with the EoS without this new degree of freedom. We use here
the $\Lambda$ hyperon as this new degree of freedom, once it is the lowest mass baryon beyond the nucleons. Nevertheless it
is { important to bear} in mind that the nature of the new degree of freedom is not relevant. Similar results can be obtained
using $\Sigma$ or $\Xi$ hyperons, or even  $\Delta$ resonances,
 by  adjusting the strength of the coupling constant for those particles.
 Even more exotic particles, as dark
 matter~\cite{Bridget,Goldman,Li,Muk} can be used without significantly
 affect the results.
 
 Using a QHD based model,   we first fix the $\Lambda$
 potential depth to its realistic value, $U_\Lambda$ = - 28 MeV in the
 traditional $\sigma\omega\rho$ parametrization, while varying the coupling constant. 
 We see that with this approach the hyperon onset reduces the maximum
 possible mass, as already expected~\cite{Vidana62,Miy88}, but the radii of the canonical
 1.4$M_\odot$ is not affected. Moreover, most of the parametrizations are unable to reproduce a 2.0$M_\odot$ neutron star. To simulate massive neutron stars, in the second approach we add a new repulsive channel, the strange vector $\phi$ meson. With the realistic $U_\Lambda$ = - 28 MeV potential, we see that although we simulate a 2.0$M_\odot$ neutron stars, and the onset of the $\Lambda$ hyperon indeed reduces the radii of the  canonical 1.4$M_\odot$, its is still not enough to explain  the compact neutron stars discussed in ref.~\cite{Hebeler,Lim,Ozel,Steiner,Steiner2}.
To accomplish this task, in the third section we also add an
attractive channel,  the strange scalar $\sigma^{*}$  meson. Since it
softs the EoS at low density it produces very compact stars. Indeed,
we are able to simulate  an EoS that predicts  1.97 $M_\odot$ as
maximum mass with a radius of 12.45 km for the canonical
1.4$M_\odot$. Furthermore, in order to give reliability to our results, we show 
that besides the realistic $U_\Lambda$ = - 28 MeV potential,  the
  hyperon-hyperon potential $U_\Lambda^\Lambda(n_0/5)$ we obtain
is weakly attractive, as expected. Our results are of the order of 
-2 MeV, a value which lies between the experimental range  
(-4 MeV to -8 MeV) proposed in ref.~\cite{Fortin95} and  -0.67 MeV
obtained in ref.~\cite{Ahn2013}.

In our last approach we generalize our results by studying the
threshold of one new
additional (not yet known) self-interacting dark matter particle, with
mass of 1.04 GeV in two different situations : we first consider that
this particle does not interact with the nucleon, corresponding to a
potential $U =$ 0 MeV and then we assume a nucleon-interacting dark matter,
with a potential depth of -33 MeV, a value close to the $\Lambda$
potential. We are able to simulate a 2.00$M_\odot$ as maximum mass and 
a canonical one with the radius of only 11.62 km.

We also compare our results with experiences of heavy-ion collisions (HIC). In ref.~\cite{Daniel}, the authors determine
the pressure of the symmetric nuclear matter up to five times nuclear
saturation density. 
They  neither  exclude the onset of
hyperons  nor of more exotic behaviour, as the quark-hadron
phase transition. We show that the emergence of a new degree of
freedom also  allows a better agreement of theory with experience.

This paper is organized as follows: in section \ref{sec2} we discuss
the QHD formalism and present the parametrization of the model {
  alongside} 
some of the physical quantities they foresee for  nuclear matter. In section \ref{sec3} we expose the results of the threshold of $\Lambda$ 
hyperon in the bulk of beta-equilibrium nuclear matter and symmetric
nuclear matter  within three different approaches, and a fourth one, studying the onset of the dark matter.
 The conclusions are  drawn in section \ref{sec4}.

\section{Formalism \label{sec2}}

The theory of the strong interacting matter is the QCD, where quarks interact with each other via the exchange of massless gauge bosons called gluons.
Since the QCD has no results for dense cold matter, an effective model is required. In this work
we use an extended version of the relativistic QHD~\cite{Serot}, whose Lagrangian density reads:

\begin{eqnarray}
\mathcal{L}_{QHD} = \sum_B \bar{\psi}_B[\gamma^\mu(i\partial_\mu  - g_{B\omega}\omega_\mu   - g_{B\rho} \frac{1}{2}\vec{\tau} \cdot \vec{\rho}_\mu)
- (m_B - g_{B\sigma}\sigma)]\psi_B  -U(\sigma) +   \nonumber   \\
  + \frac{1}{2}(\partial_\mu \sigma \partial^\mu \sigma - m_s^2\sigma^2) - \frac{1}{4}\Omega^{\mu \nu}\Omega_{\mu \nu} + \frac{1}{2} m_v^2 \omega_\mu \omega^\mu  
 + \frac{1}{2} m_\rho^2 \vec{\rho}_\mu \cdot \vec{\rho}^{ \; \mu} - \frac{1}{4}\bf{P}^{\mu \nu} \cdot \bf{P}_{\mu \nu}  , \label{s1} 
\end{eqnarray}
in natural units. 
 $\psi_B$  are the baryonic  Dirac fields, which can be the nucleons, or a new degree of freedom, in this case, $\Lambda$ hyperon. The $\sigma$, $\omega_\mu$
and $\vec{\rho}_\mu$ are the mesonic fields.
 The $g's$ are the Yukawa coupling constants that simulate the strong
 interaction, $m_B$ is the mass of the baryon $B$ and $m_s$, $m_v$,  and $m_\rho$ are
 the masses of the $\sigma$, $\omega$, and $\rho$ mesons respectively.
 The antisymmetric mesonic field strength tensors are given by their usual expressions as presented in~\cite{Glen}.
  The $U(\sigma)$ is the self-interaction term introduced in ref.~\cite{Boguta} to reproduce some of the saturation properties of the nuclear matter and is given by:
 
 \begin{equation}
U(\sigma) =  \frac{1}{3!}\kappa \sigma^3 + \frac{1}{4!}\lambda \sigma^{4} \label{s2} .
\end{equation}
 
 Finally, $\vec{\tau}$ are the Pauli matrices. In order to describe a neutral, chemically stable  matter, we add leptons as free Fermi gases:
 
 \begin{equation}
 \mathcal{L}_{lep} = \sum_l \bar{\psi}_l [i\gamma^\mu\partial_\mu -m_l]\psi_l , \label{s3}
 \end{equation}
 where the sum runs over the two lightest leptons ($e$ and $\mu$).

The mesonic fields are obtained via mean field approximation (MFA)~\cite{Serot,Glen,Lopes2012} and the EoS by thermodynamic relations~\cite{Glen,Greiner}.

To describe the properties of  nuclear matter we use a slightly modified version of the well-known GM1 parametrization~\cite{Glen2},
a widely accepted parametrization~\cite{Lopes14,Glen,Lopes2012,Benito2,Paoli,Weiss1,Weiss2,Lopes2013,Lopes2015} that is able to reasonably describe both, nuclear matter and stellar structure, consistent with experimental and astrophysical observations~\cite{Lopes2013}. 
In this work we just reduce the strength of the $\rho$ coupling, reducing the symmetry energy slope $L$ from 94 MeV to 87.9 MeV~\cite{Lopes14}, a value closer
to what is inferred in recent
observations~\cite{Lim,Steiner,Steiner2}. 
 This slope can be reduced even further,
  \cite{Rafa2011,alex2012}, but for the purpose of the present work,
  the modifications introduced in \cite{Lopes14} suffice.

In Table~\ref{T1} we show the parameters of the model and its previsions for five nuclear matter properties at saturation density: saturation density
point ($n_0$), incompressibility ($K$), binding energy per baryon ($B/A$), symmetry energy ($S_0$) and its slope ($L$). 

\begin{table}[ht]
\begin{center}
\begin{tabular}{|c|c||c|c|c|}
\hline
 \multicolumn{2}{|c||}{ Parameters} & \multicolumn{2}{c|}{Previsions at $n_0$} \\

 \hline
 $(g_{N\omega}/m_v)^2$   &  7.148 $fm^2$  & $n_0~(fm^{-3})$  &  0.153     \\
 \hline
  $(g_{N\sigma}/m_s)^2$ & 11.785 $fm^2$  & $K$ (MeV) & 300    \\
 \hline
   $(g_{N\rho}/m_\rho)^2$ & 3.880 $fm^2$  & $B/A$ (MeV) & -16.3     \\
 \hline
  $\kappa/M_N$  & 0.005894 & $S_0$ (MeV)& 30.5  \\
\hline
 $\lambda$ & -0.006426 & $L$ (MeV) & 87.9    \\
\hline

\end{tabular} 
\caption{Slightly modified GM1 parametrization. Parameters of the model and previsions.} 
\label{T1}
\end{center}
\end{table}

\section{Results \label{sec3}}

To study the effects of the onset of a new degree of freedom in detail, we divide this section in
four different approaches.  The possible existence of hyperons in the
interior of neutron stars is an old but
very active field as can be seen by the large number of studies published in the last five years~\cite{Weiss1,Weiss2,Lopes2013,Lopes2015,
alex2012,Massot2012,Katayama2012,Drago2014,Gomes2014,BB2014,Ortel2015,Vidana2016}. In this work we consider the $\Lambda$ hyperon only
because the presence of  other
strange  particles would make this study strongly model and parameter dependent.

\subsection{Role of the coupling constant}

\begin{figure}[ht] 
\begin{centering}
 \includegraphics[angle=270,
width=0.70\textwidth]{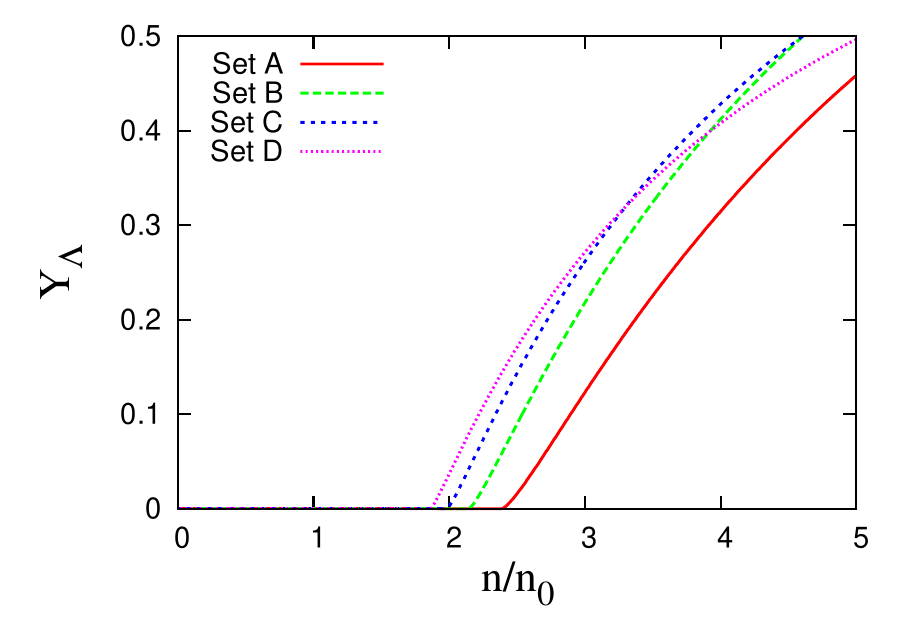}
\caption{(Color online) $\Lambda$ threshold and population for a fixed potential depth $U_\Lambda$ = -28 MeV.} \label{F1}
\end{centering}
\end{figure}

The basic constituents of neutron stars are neutrons and protons in $\beta$-equilibrium. Since
both are fermions, as the baryon density increases, so do the Fermi
momentum and the Fermi energy, according to the Pauli principle.
Ultimately, the Fermi energy exceeds the masses of the heavier baryons. 
 The influence of the hyperons in the EoS and neutron star mass-radius relation,
strongly depend of the coupling constant of the hyperons with the mesonic field. 
It is well known that the $\Lambda$ potential depth $U_\Lambda$ = - 28 MeV~\cite{Glen2}, so the coupling constants
for $\omega$ and $\sigma$ meson cannot be varied independently. 
So, in this section we fix the potential depth and vary the values of
the coupling constants  and see
how they affect the hyperon onset and population, the EoS, and 
  consequently the mass-radius relation
of the neutron stars. Similar works are found in the literature~\cite{Glen,Glen2,Vidana62,Miy88},
however, they just analyze the effects on the maximum mass. Let's
study, in more detail, the influence on the radii of the neutron stars
as well. In Table~\ref{T2} we show  four sets of  values for the
coupling constants utilized in this section.  We also plot the
  corresponding threshold density for the $\Lambda$ (what usually
  corresponds to
$Y_\Lambda$ around $10^{-4}$), and the density at which  $Y_\Lambda$ reaches 0.1. In Fig.~\ref{F1} we plot the $\Lambda$ fraction
$Y_\Lambda = n_\Lambda/n$ with four sets for densities up to five times the saturation one.

\begin{table}[ht]
\begin{center}
\begin{tabular}{|c||c|c|c|c|c|}
\hline 
 Set &  $g_{\Lambda,\omega}/g_{N,\omega}$ &  $g_{\Lambda,\sigma}/g_{N,\sigma}$  &  $U_\Lambda$ (MeV) & $n/n_0$ ($Y_\Lambda = 10^{-4})$
& $n/n_0$ ($Y_\Lambda = 10^{-1})$ \\
  \hline
  A & 0.75 & 0.674 & -28 & 2.40 & 2.90 \\
 \hline
  B & 0.50 & 0.483 & -28 & 2.15 & 2.54\\
 \hline
  C & 0.25 & 0.291 & -28 & 1.99 & 2.34\\
 \hline
 D & 0.00  & 0.099 & -28  & 1.83 & 2.22\\
 \hline

\end{tabular} 
\caption{Different sets for $\Lambda$-mesons coupling constants with a fixed potential depth  $U_\Lambda$ = -28 MeV and their threshold densities.} 
\label{T2}
\end{center}
\end{table}

We see that the higher the value of $g_{\Lambda,\omega}$, the higher
the density of the hyperon threshold, varying from 2.40 times the nuclear 
saturation density for set A
to 1.83 times  for set D.  On the other hand, for a small increase 
in the total density ( $\Delta$ n varying from 0.35 $n_0$ to 0.50 $n_0$), $Y_\Lambda$ increases
three orders of magnitude, reaching 10 percent.
 We also can see that the curves  crosses each other. This is due to the Pauli
blocking of the hyperons, as explained in ref.~\cite{Lopes2013} for the strangeness
fraction. Once there are more hyperons
in set D at low densities, the onset of new $\Lambda$ particles is
more energetically  favorable, for instance, in set A
because the Fermi sea for low Fermi moment in set D is already filled.

\begin{figure}[ht] 
\begin{centering}
 \includegraphics[angle=270,
width=0.70\textwidth]{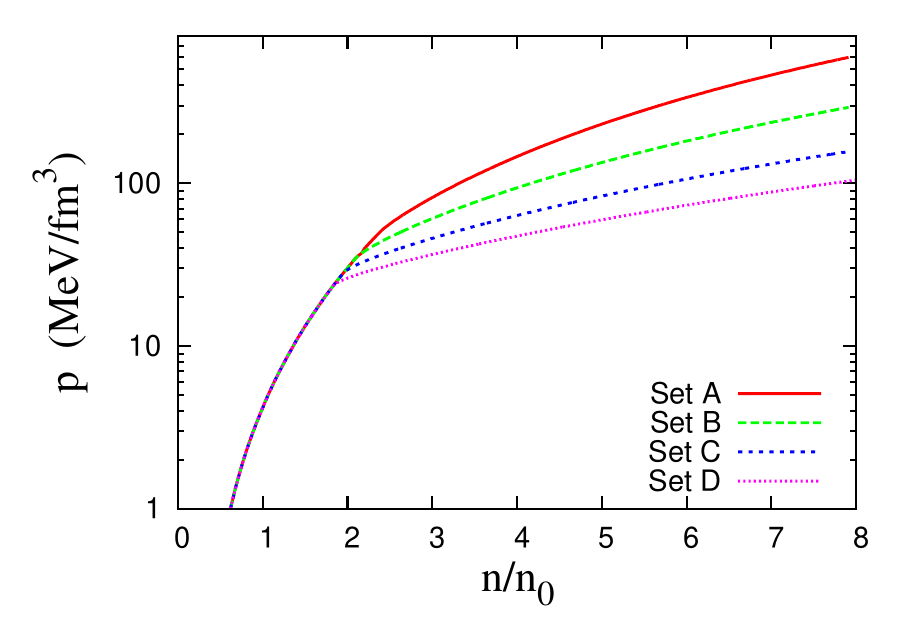}
\caption{(Color online) EoS for different sets with a fixed potential depth  $U_\Lambda$ = -28 MeV.} \label{F2}
\end{centering}
\end{figure}

Now we plot in Fig.~\ref{F2} the EoS for the four sets presented in Table~\ref{T2}. We see that the higher the value of 
$g_{\Lambda,\omega}$, the stiffer the EoS. Unlike Fig.~\ref{F1} there
is no crossing of the curves. This is due to the fact 
that the main term of the pressure is the vector meson $\omega$ and it is proportional to the density $n$. So, indeed, the
curves deviate from  each other as the density increases.

\begin{table}[ht]
\begin{center}
\begin{tabular}{|c||c|c|c|c|c|}
\hline 
 Set &  $ M_{\max}/M_\odot$  &  $R_{M_{\max}}$ (km)& $n/n_0 ~(M_{max})$ & $R_{1.4M_\odot}$ (km) & $n/n_0 ~M(1.4M_\odot)$ \\
 \hline
 No hyperons & 2.38 & 11.90 & 5.50 & 13.72 & 2.18 \\
 \hline
 A & 2.07  & 12.11 & 5.88 & 13.72 & 2.18 \\
 \hline
  B & 1.71 & 12.68 & 5.64 & 13.72 & 2.19 \\
 \hline
  C & 1.44 & 13.24 & 4.18 & 13.62 & 2.80 \\
 \hline
  D & 1.28 & 13.44 & 3.66 & - & -  \\
 \hline
\end{tabular} 
\caption{ Some stellar properties  for different $\Lambda$-mesons coupling constants with a fixed potential depth  $U_\Lambda$ = -28 MeV.} 
\label{T3}
\end{center}
\end{table}

\begin{figure*}[ht]
\begin{tabular}{cc}
\includegraphics[width=6.0cm,height=7.0cm,angle=270]{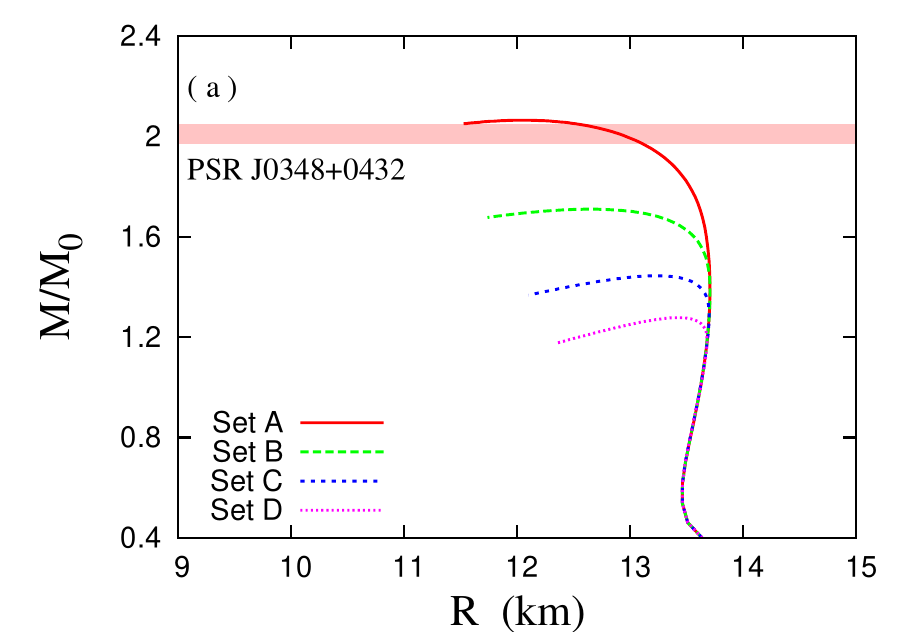} &
\includegraphics[width=6.0cm,height=7.0cm,angle=270]{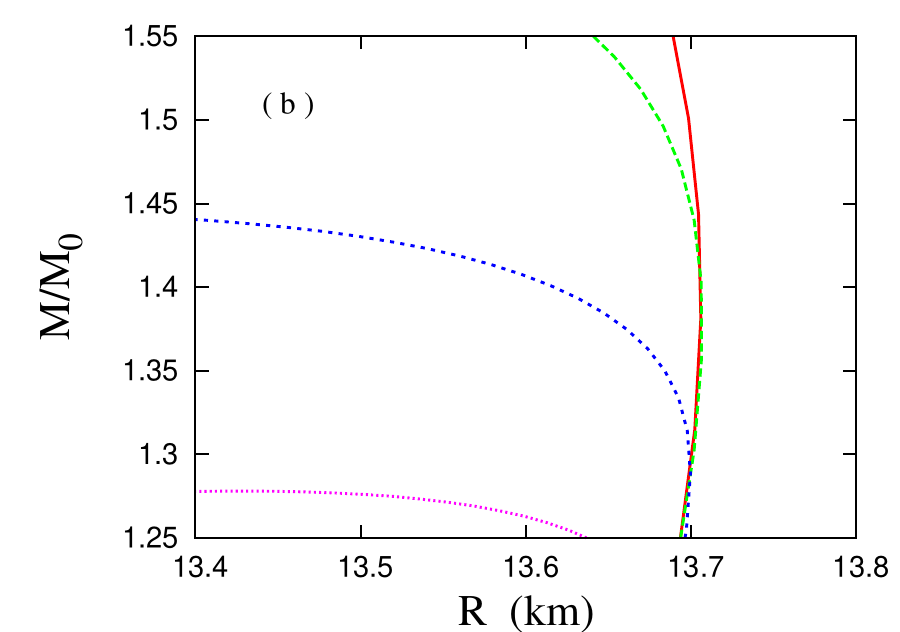} \\
\end{tabular}
\caption{(Color online) (a) Mass-radius relation obtained via TOV
  solution, the hatched area comprises the uncertainty 
about the mass of the PSR J0348+0432.  (b) Zoom in the mass around $1.4M_\odot$.} \label{F3}
\end{figure*}

To conclude this section we solve the TOV structural equations~\cite{TOV} and plot the mass-radius relation in Fig.~\ref{F3}.
Here, and in the rest of this work we use the BPS~\cite{BPS} equation to simulate the neutron star crust.
The most massive pulsar yet known is the PSR J0348+0432~\cite{Antoniadis} with a mass of $2.01~\pm~0.04~M_\odot$. So, for
an EoS to be valid, it needs to explain this massive neutron star. Another constraint is the radius of the canonical
1.4$M_\odot$ pulsar. According to ref.~\cite{Hebeler,Lim,Ozel,Steiner,Steiner2} the maximum radius for these stars
lie between 12.50 to 13.9 km. We plot the main results for the TOV solutions in Table~\ref{T3}.

 The maximum neutron star mass before the hyperon onset varies
  from 1.56$M_\odot$ for set A to 1.12$M_\odot$ for set
D. Also, from the central density of the 1.4$M_\odot$, we can see that
only set C produces more than 10$\%$ of hyperons in its core.

By looking at Fig.~\ref{F3}, we can see that according to ref.~\cite{Antoniadis}, only set A is valid as
EoS of dense matter. The radius of the canonical 1.4$M_\odot$ is 13.72 km for set A and B (and also if hyperons are not present)
and slightly lower for set C. This is due to the onset of the $\Lambda$ particle at low density in set C.
A new degree of freedom as the $\Lambda$ hyperon not only reduces the maximum mass, but compresses
the neutron star. When the hyperon fraction becomes relevant, there is a ``turn to the left" in the mass-radius relation, compressing the subsequent
neutron stars, reducing { their} radii. For a hyperon potential
depth of -28 MeV, the threshold of hyperons appears too late
 for this ``turn to the left"  affect
the canonical  1.4$M_\odot$ to values of radii that agree with the ref.~\cite{Hebeler,Lim,Ozel,Steiner,Steiner2}.
But for more massive neutron stars, the large amount of hyperons compresses significantly the star.
For instance, if someone wonders about the radius of a
2.01$M_\odot$, the PSR J0348+0432, 
the answer is  13.10 km, if
we assume set A as the best parametrization, or 13.48 km if we believe that there is no hyperon,
 because the ``turn to the left" happens earlier in set A.
If we want to compress the canonical 1.4$M_\odot$ neutron star, the
$\Lambda$ threshold needs to  take place at earlier densities, forcing us to choose set D.
However to explain the massive  PSR J0348+0432, we need to find a way to stiff the EoS at
high densities to simultaneous simulate a massive and a compact neutron star.

\subsection{Role of the strange vector meson $\phi$}

We see that the emergence of a new degree of freedom causes a ``turn to the left" in the mass-radius relation, compressing the subsequent stars.
The earlier  the onset of this new particle, the lower  the mass that is affected by this ``turn to the left". But we also see that the threshold 
of a new particle softens the EoS. To stiffen the EoS, we need to increase the repulsive channel. The term responsible to this in eq.~(\ref{s1})
is the $\Lambda-\omega$ coupling constant. However, the increase of the $\Lambda-\omega$ channel  will affect the $U_\Lambda$ potential, and 
the ``turn to the left" will not affect the canonical 1.4$M_\odot$. One way to overcome this difficulty is  to add in the Lagrangian a new 
repulsive channel. We use here here the strange vector $\phi$ meson~\cite{Weiss1,Weiss2,Lopes2013}:

\begin{equation}
\mathcal{L}_{YY\phi} =  -g_{Y,\phi}\bar{\psi}_Y(\gamma^\mu \phi_\mu)\psi_Y + \frac{1}{2} m_\phi^2 \phi_\mu \phi^\mu
 - \frac{1}{4}\Phi^{\mu \nu}\Phi_{\mu \nu}   . \label{s4} 
\end{equation}

\begin{table}[ht]
\begin{center}
\begin{tabular}{|c||c|c|c|c|c|c|}
\hline 
 Set &  $g_{\Lambda,\omega}/g_{N,\omega}$ &  $g_{\Lambda,\sigma}/g_{N,\sigma}$  & $g_{\Lambda,\phi}/g_{N,\omega}$ &
 $U_\Lambda$ (MeV) & $n/n_0$ ($Y_\Lambda = 10^{-4})$ & $n/n_0$ ($Y_\Lambda = 10^{-1})$ \\
   \hline
  D-I & 0.00 & 0.099 &0.85 &-28 & 1.83 & 2.37 \\
 \hline
  D-II & 0.00 & 0.099 &1.25 &-28 & 1.83 & 2.56\\
 \hline
  D-III & 0.00 & 0.099 &1.55 &-28 & 1.83 & 2.79\\
 \hline
\end{tabular} 
\caption{Different sets for $g_{\Lambda,\phi}/g_{N,\omega}$  with a fixed $U_\Lambda$ = -28 MeV and their respective $\Lambda$ threshold.} 
\label{T5}
\end{center}
\end{table}

\begin{figure}[ht] 
\begin{centering}
 \includegraphics[angle=270,
width=0.70\textwidth]{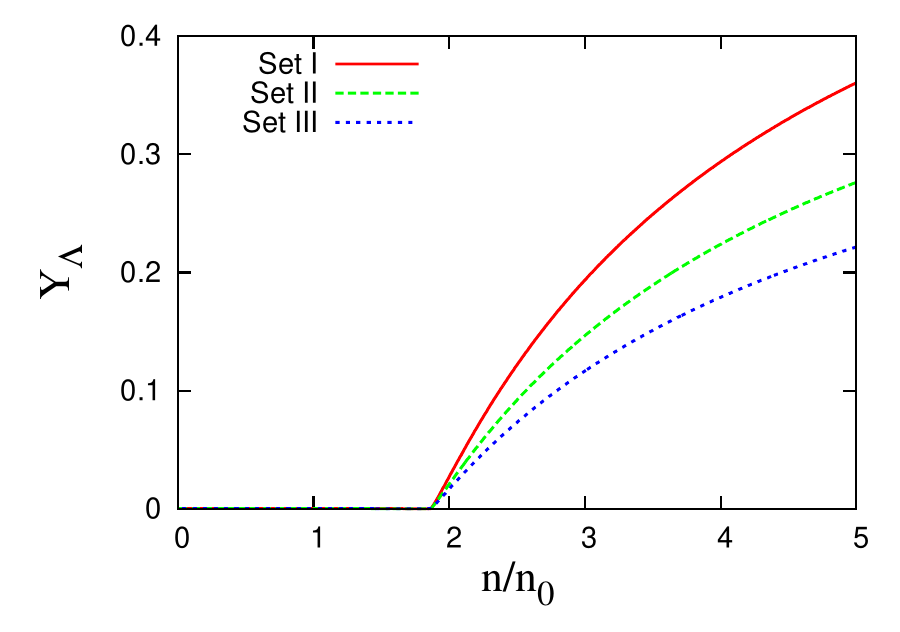}
\caption{(Color online) $\Lambda$ threshold and population for different values of $\Lambda-\phi$ coupling, with a fixed $U_\Lambda$ = -28 MeV.} \label{F7}
\end{centering}
\end{figure}

The $\phi$ field is analogous to the $\omega$ field and its expected value is also obtained via MFA~\cite{Serot,Glen,Lopes2012}.
Adding a new mesonic field has two advantages. First, since the $\phi$ does not couple to the nucleon, it does not affect the properties
of the nuclear matter. Second, since the $\phi$ field is zero below the hyperon threshold density, it does not affect the potential depth,
and has little influence on the point of the ``turn to the left".  Now we fix $\Lambda-\omega$ and $\Lambda-\sigma$ couplings as in set D,
once this set produces the earlier ``turn to the left" .
Indeed, all sets in this section are derived from set D of
Table~\ref{T2}, only varying the $\Lambda-\phi$ coupling constant. So,
we numbered the sets from D-I to D-III. In the 
figures the D is omitted to avoid burdening the notation.
The values utilized in this calculation are presented in Table~\ref{T5}.

\begin{figure}[ht] 
\begin{centering}
 \includegraphics[angle=270,
width=0.70\textwidth]{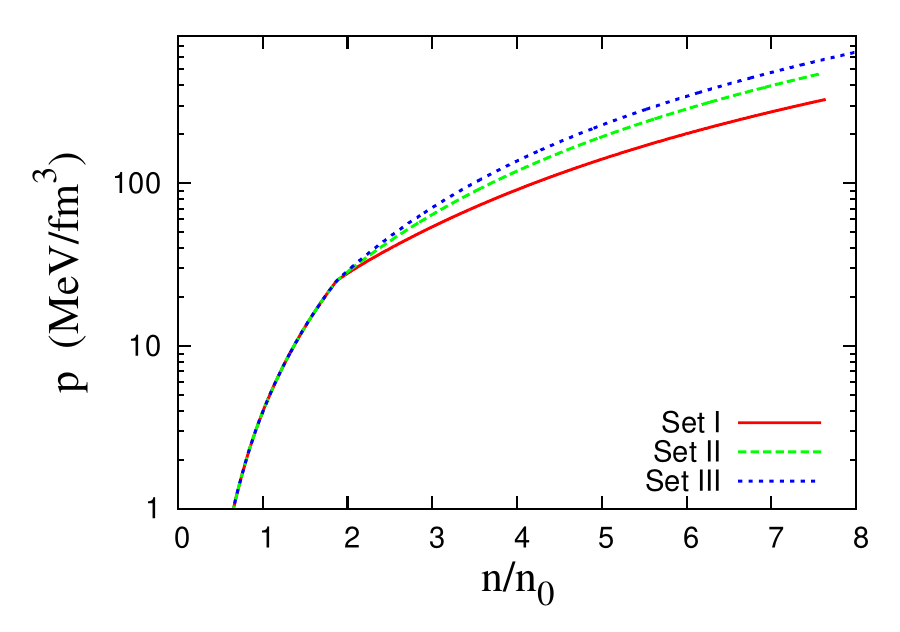}
\caption{(Color online) EoS for different values of $\Lambda-\phi$ coupling, with a fixed $U_\Lambda$ = -28 MeV.} \label{F8}
\end{centering}
\end{figure}

\begin{figure*}[ht]
\begin{tabular}{cc}
\includegraphics[width=6.0cm,height=7.0cm,angle=270]{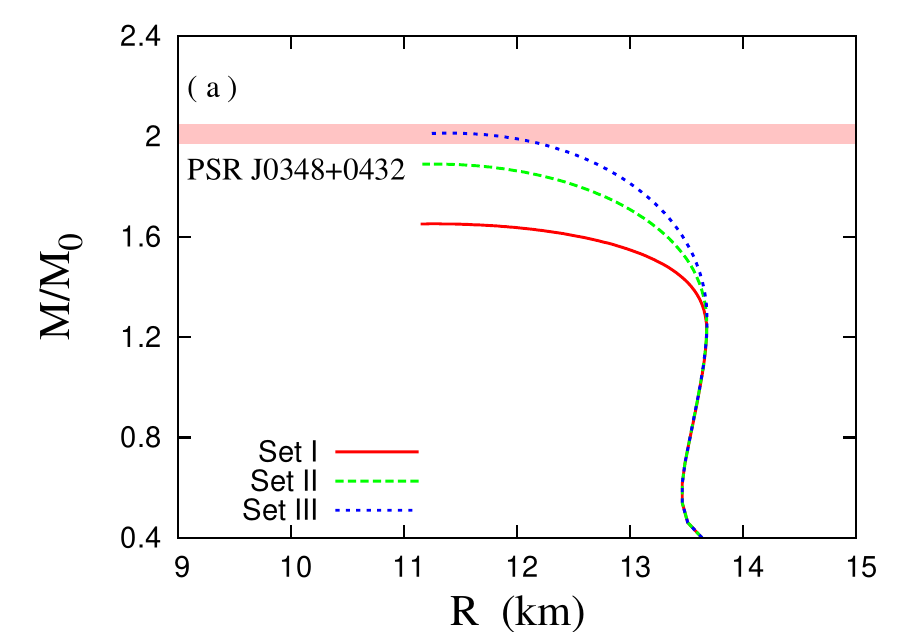} &
\includegraphics[width=6.0cm,height=7.0cm,angle=270]{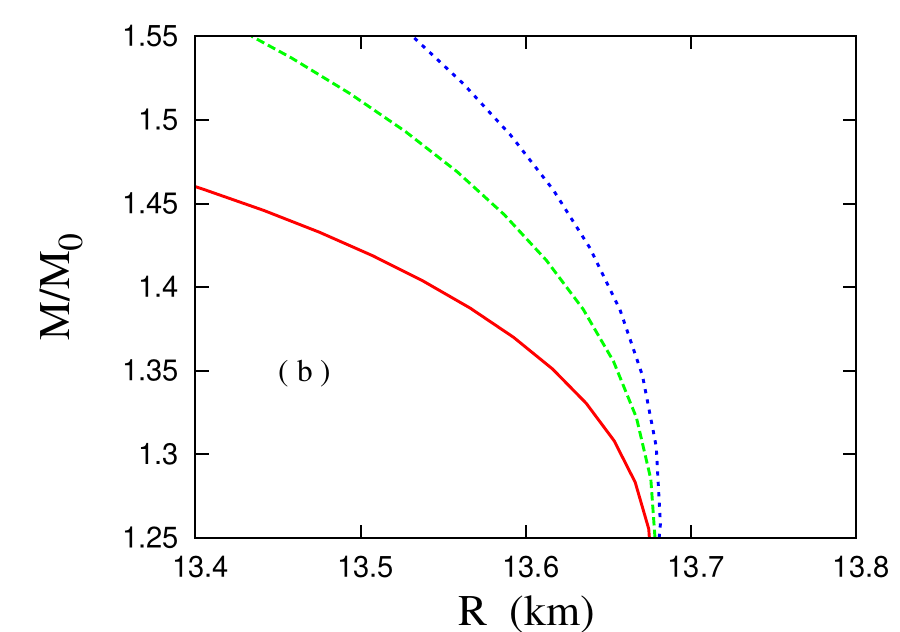} \\
\end{tabular}
\caption{(Color online) (a) Mass-radius relation obtained via TOV solution with the inclusion of the $\phi$ meson. 
The hatched area comprises the uncertainty
about the mass of the PSR J0348+0432.  (b) Zoom in the mass around $1.4M_\odot$.} \label{F9}
\end{figure*}

Now we plot the $\Lambda$ fraction in Fig.~\ref{F7}. As we can see, since $\phi$ field is zero before the hyperon threshold, and the potential depth
are the same for all sets, the hyperon onset is the same for all values of $\Lambda-\phi$ coupling, around 1.83 $n_0$.
Since $\phi$ is a repulsive channel, the
{ stronger}  the coupling, the stronger  the hyperon suppression at
high densities.  Within these models, the $\phi$ field supresses
  the hyperon creation, and as can be seen from Table~\ref{T5}, the stronger
the $\Lambda-\phi$ coupling, the higher the density where $Y_\Lambda$
reaches 0.1. As a consequence, $Y_\Lambda$ never surpasses 0.3 
in set D-III.
 Also, the $\phi$ field contributes in the pressure,
stiffening the EoS as we can see in Fig.~\ref{F8}.

\begin{table}[ht]
\begin{center}
\begin{tabular}{|c||c|c|c|c|c|}
\hline 
 Set &  $ M_{\max}/M_\odot$  &  $R_{M_{\max}}$ (km) & $n/n_0$ $(M_{max})$& $R_{1.4M_\odot}$ (km) & $n/n_0$ $M(1.4M_\odot)$ \\
 \hline
 D-I & 1.65  & 11.42 & 7.26 & 13.54 & 2.70  \\
 \hline
  D-II & 1.89 & 11.64 & 6.94 & 13.62 & 2.42 \\
 \hline
  D-III & 2.01 & 11.87 & 6.70 & 13.65 & 2.33\\
 \hline
\end{tabular} 
\caption{Main properties of neutron stars with the inclusion of the $\phi$ meson for $U_\Lambda$ = -28 MeV.} 
\label{T6}
\end{center}
\end{table}

Now we plot the mass-radius relation in Fig.~\ref{F9} with the main
properties resumed in Table~\ref{T6}.  As all models are derived
  from set D, the maximum  stellar mass without hyperons 
in the core is 1.12$M_\odot$. We can see that the $\phi$ meson
is able to stiff the EoS enough to produce neutron stars as massive as
the PSR J0348+0432. However, although the ```turn to the left" happens
very earlier, affecting neutron stars with low masses due to the onset
of hyperons at low densities, the EoS becomes stiff too soon, and the
canonical 1.4$M_\odot$ is not sufficiently compressed to be in
agreement with the results presented in
refs.~\cite{Hebeler,Lim,Ozel,Steiner,Steiner2}.  As a repulsive
  channel, the higher the $\phi$ field, the lower the central 
density of the canonical 1.4$M_\odot$ star.
Also,the price we pay to produce massive neutron stars with the help of the
$\phi$ meson is to obtain a canonical mass star with a large
radius. Nonetheless, there is a small reduction of the radius from 13.72 km to 13.65 km  in set D-III, the only which is able to 
reproduce a 2$M_\odot$ star.
 We conclude this section stating that the $\phi$ meson alone is not
 able to simultaneous produces massive
and compact neutron stars. We need to find a way to keep the EoS stiff at high densities, but make it soft at low ones.

\subsection{Role of the strange scalar meson $\sigma^{*}$}

Now we give one more step and add a new strange meson, the scalar
$\sigma^{*}$. As the vector $\phi$ meson, the scalar one
just { couples} to the $\Lambda$ particles, and does not affect any of the properties of nuclear matter presented in Table~\ref{T1}.
Also, the $\sigma^{*}$ field is zero in the absence of $\Lambda$ particles, therefore
 it would not affect the hyperon threshold.  The Lagrangian of $\sigma^{*}$ is analogous to the $\sigma$ and reads:

\begin{equation}
\mathcal{L}_{YY\sigma^{*}} =   g_{Y,\sigma^{*}}(\bar{\psi}_Y \psi_Y )\sigma^{*} + \frac{1}{2} \bigg (\partial^\mu\sigma^{*}\partial_\mu\sigma^{*} -
 m_{\sigma^*}^2 \sigma^{*2} \bigg ) . \label{s6} 
\end{equation}

\begin{table}[ht]
\begin{center}
\begin{tabular}{|c||c|c|c|c|c|c|}
\hline 
 Set &   $g_{\Lambda,\sigma^{*}}/g_{N,\sigma}$  & $g_{\Lambda,\phi}/g_{N,\omega}$ &$U_\Lambda$ (MeV) &  $U_\Lambda^\Lambda(n_0/5)$
& $n/n_0$ ($Y_\Lambda = 10^{-4})$ & $n/n_0$ ($Y_\Lambda = 10^{-1})$ \\
  \hline
  D-IV & 2.96 & 2.45  &-28 & -2.2 MeV & 1.83 & 2.07\\
 \hline
  D-V & 2.90 & 2.39  &-28 & -2.1 MeV & 1.83 & 2.05 \\
 \hline
  D-VI & 2.82 & 2.32  &-28 & -2.0 MeV & 1.83 & 2.04 \\
 \hline

\end{tabular} 
\caption{Different sets for $g_{\Lambda,\sigma^{*}}/g_{N,\sigma}$ and $g_{\Lambda,\phi}/g_{N,\omega}$  with a fixed
 $U_\Lambda$ = -28 MeV and  the calculated value for the $ U_\Lambda^\Lambda(n_0/5)$.} 
\label{T7}
\end{center}
\end{table}

\begin{figure}[ht] 
\begin{centering}
 \includegraphics[angle=270,
width=0.70\textwidth]{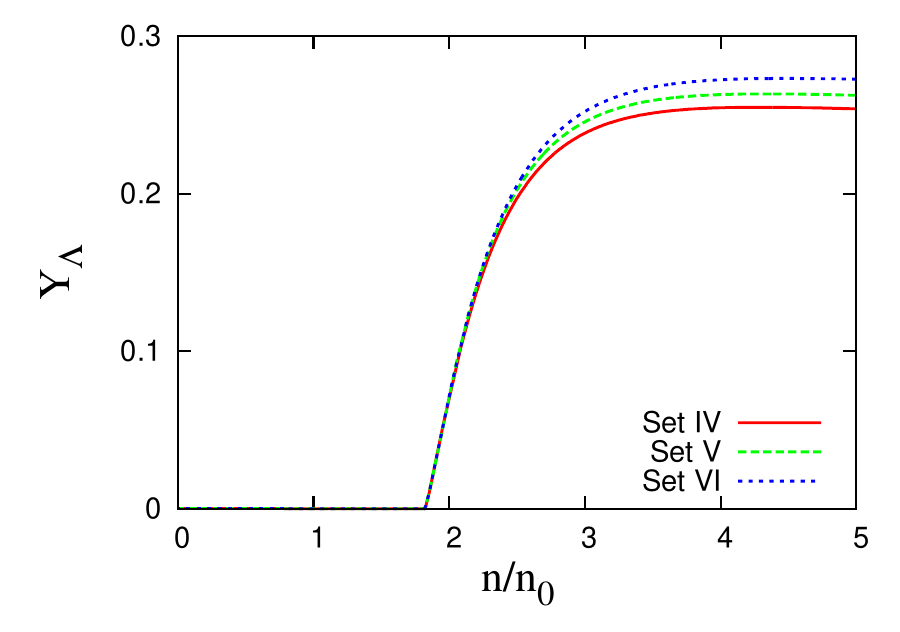}
\caption{(Color online) $\Lambda$ threshold and population for different values of $\Lambda-\sigma^{*}$ and
 $\Lambda-\phi$ coupling.} \label{F11}
\end{centering}
\end{figure}

The $\sigma^{*}$ is an attractive scalar meson, while the $\phi$ is a
repulsive vector meson. Therefore, the $\sigma^{*}$  dominates at low densities,
softening the EoS and increasing the amount of hyperons. At high
densities, the $\phi$ meson dominates, causing a strong suppression of the
hyperons and stiffening the EoS. This behavior is analogous to the
$\delta-\rho$ competition 
in the symmetry energy reported in ref.~\cite{Lopes14}.
 In Table~\ref{T7} we show the values of the coupling constants
for the $\Lambda-\sigma^{*}$ and $\Lambda-\phi$ mesons. Again, the values are derived from set D of Table~\ref{T2}.  All these
values have been chosen  in such a way that the EoS is able to reproduce the PSR J0348+0432, i.e, their maximum masses 
are in the range 2.01 $\pm$ 0.04$M_\odot$. Moreover, besides the
reproduction of the well known $U_\Lambda$ = -28 MeV,  we have
  calculated the  hyperon-hyperon potential $U_\Lambda^\Lambda(n_0/5)$
  and obtained a weakly attractive potential. Our results are of the order of 
-2 MeV, a value which lies between the experimental range  
(-4 MeV to -8 MeV) proposed in ref.~\cite{Fortin95} and  -0.67 MeV
obtained in ref.~\cite{Ahn2013}.
As we can see from Table~\ref{T7} all models yield similar results.
All produce the hyperon threshold at 1.83 $n_0$, and even similar densities where $Y_\Lambda$ reaches 0.1.

Now we plot in Fig.~\ref{F11} the hyperon fraction for the sets of Table~\ref{T7}. We see that in all cases, the $\Lambda$ fraction grows quickly due to the very 
larger $\Lambda-\sigma^*$ coupling.  Indeed, when compared with the original set D, we can see that the density where
$Y_\Lambda$ reaches 0.1 dropped from 2.22 $n_0$ for values around 2.05 $n_0$. As the densities increase and the lower levels of the Fermi see are
filled, the $\phi$ mesons act to suppress the $Y_\Lambda$ at high
densities, resulting in all cases in a low $Y_\Lambda$, which never surpasses 0.3 at high densities.

\begin{figure}[ht] 
\begin{centering}
 \includegraphics[angle=270,
width=0.70\textwidth]{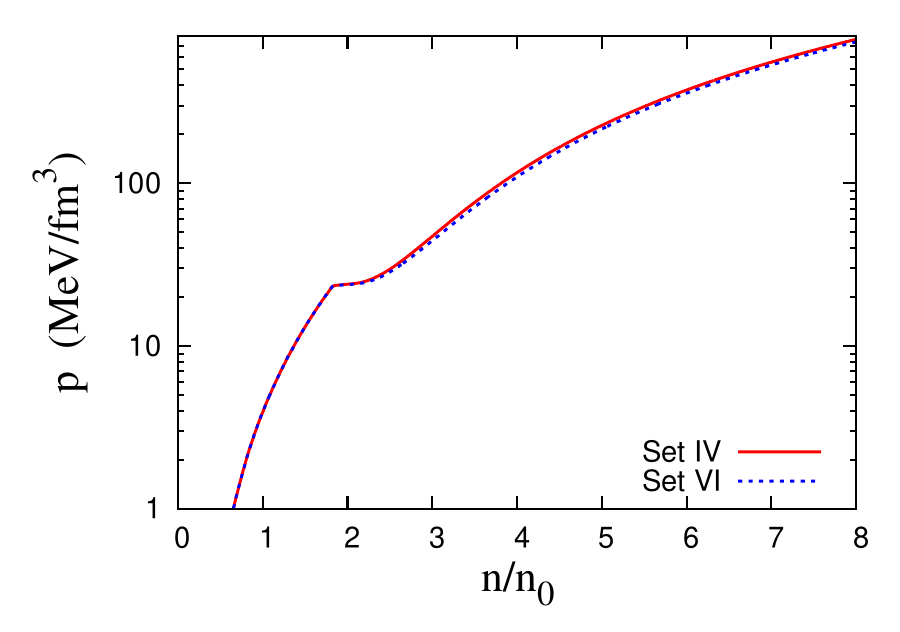}
\caption{(Color online) EoS for sets D-IV and D-VI.} \label{F12}
\end{centering}
\end{figure}

\begin{figure*}[ht]
\begin{tabular}{cc}
\includegraphics[width=6.0cm,height=7.0cm,angle=270]{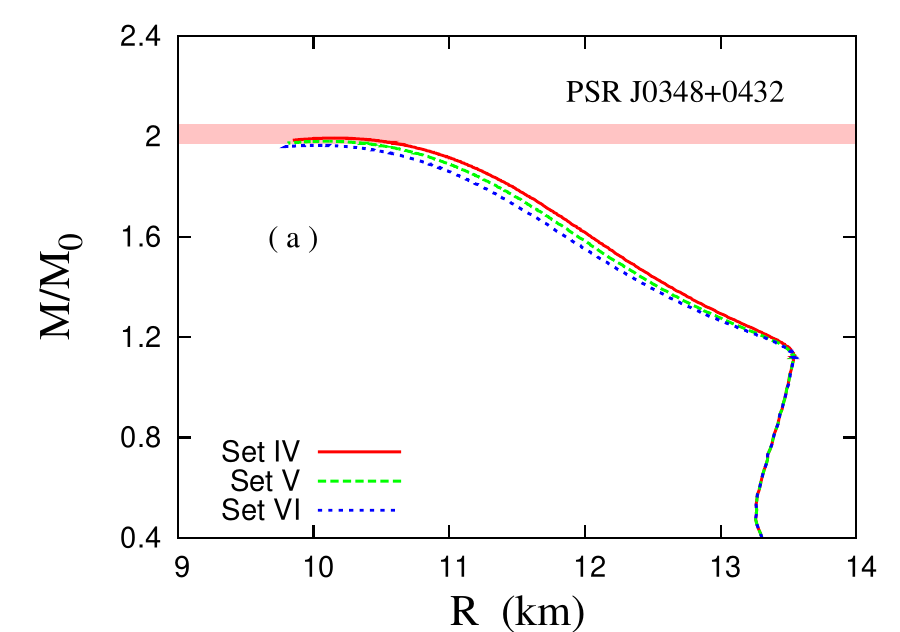} &
\includegraphics[width=6.0cm,height=7.0cm,angle=270]{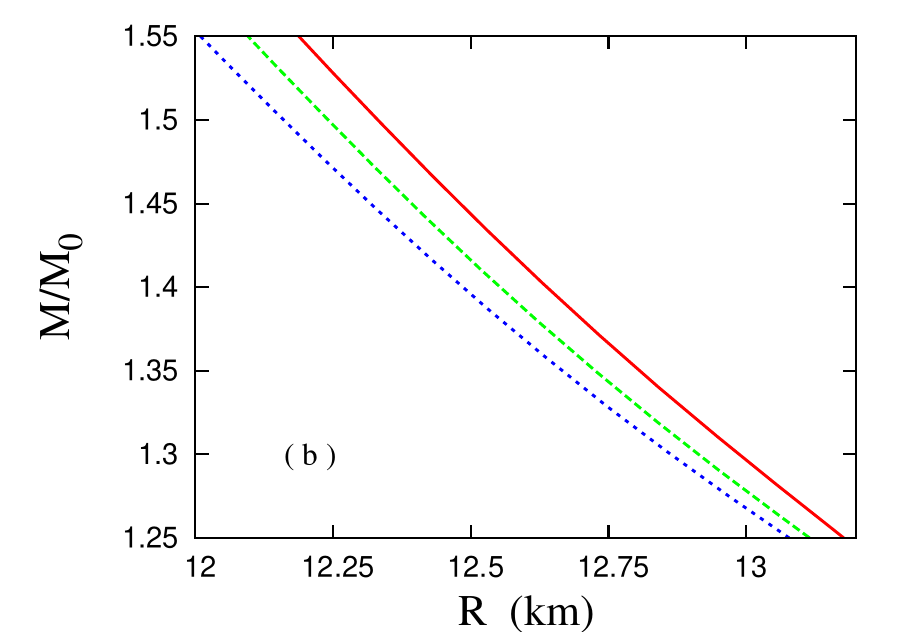} \\
\end{tabular}
\caption{(Color online) a) Mass-radius relation considering both  $\sigma^{*}$ and $\phi$ mesons constrained to a maximum mass 
in agreement with the PSR J0348+0432 pulsar. b) Zoom in the mass around $1.4M_\odot$.} \label{F13}
\end{figure*}

The differences in the EoS are even smaller. Since all the EoS are
adjusted to reproduce the PSR J0348+0432 as maximum mass, they are all
very similar. Hence, we just plot  sets D-IV and  D-VI, that bear the
weakest and strongest $\Lambda-\phi$ coupling constant.
Of course we could increase the repulsive channel and simulate even
higher mass neutron stars, however this would imperatively
increase the radius of the canonical 1.4$M_\odot$. In the same way we cannot arbitrary vary the $\Lambda-\sigma^{*}$ coupling constant.
Increasing both  $\Lambda-\phi$ and  $\Lambda-\sigma^{*}$ coupling constant would create a region of thermo-mechanical instability, where the
Le Chatelier principle would be violated~\cite{Glen}. On the other hand, reducing both coupling constants would reduce the maximum mass to values 
below the limit of the PSR J0348+0432 pulsar.
The results are presented in Fig.~\ref{F12}. As expected, these two sets produce very similar EoS.

\begin{table}[ht]
\begin{center}
\begin{tabular}{|c||c|c|c|c|c|}
\hline 
 Set &  $ M_{\max}/M_\odot$  &  $R_{M_{\max}}$ (km) & $n/n_0$ $(M_{max})$ & $R_{1.4M_\odot}$ (km) & $n/n_0$ $M(1.4M_\odot)$ \\
 \hline
 D-IV & 1.99  & 10.19 & 7.77 & 12.61 & 3.45  \\
 \hline
  D-V & 1.98 & 10.08 & 8.07 & 12.52 & 3.53\\
 \hline
  D-VI & 1.97 & 10.03 & 8.11 & 12.45 & 3.63\\
 \hline
\end{tabular} 
\caption{Main properties of neutron stars with the inclusion of the $\sigma^{*}$ and $\phi$ mesons for $U_\Lambda$ = -28 MeV.} 
\label{T8}
\end{center}
\end{table}

Now, lets solve the TOV equations and obtain the mass-radius
relation. The results are plotted in Fig~\ref{F13} and the main
relevant properties are resumed in Table~\ref{T8}.  Again, the
  maximum mass for which the star has no hyperons in its core is 1.12$M_\odot$. For more massive stars, the presence of the
$\sigma^{*}$ softens the EoS at low densities. This behavior causes the ``turn to the left" in mass-radius relation to be more pronounced, compressing the subsequent neutron
stars and producing a significantly lower value for the radius
of the 1.4$M_\odot$ star.  This  also makes the central densities
  of the canonical mass being around 3.5 $n_0$, which 
that if the hyperons are the cause of the small radii, their population must be above 20$\%$ in the core of the 1.4$M_\odot$.

The strong   $\Lambda-\sigma^{*}$ coupling
makes the EoS very soft at low densities, which produces a
very small radius to the canonical 1.4$M_\odot$. Still, the
$\Lambda-\phi$ coupling is strong enough to produce a stiff EoS at high density,
able to reproduce a 2.0$M_\odot$ maximum mass neutron star.  It is
  worth bearing in mind that the value  12.45 km for the
radius of canonical mass is very close to the limit of the GM1 model
for the realistic $U_\Lambda$ = -28 MeV,
 once, as explained earlier, changing the coupling constants would
 either violate the Le Chatelier principle~\cite{Glen} or reduce the
 maximum mass to below the experimental limit.

It is important to empathize that with the help of $\sigma^{*}$ meson
we are able to simulate neutron stars as massive as the
PSR J0348+0432~\cite{Antoniadis}, and radii in the range proposed in
ref.~\cite{Hebeler,Lim,Ozel,Steiner,Steiner2}, with reasonable
$U_\Lambda$ and $U_\Lambda^\Lambda (n_0/5)$
potentials~\cite{Glen2,Fortin95,Ahn2013}. 
Comparing the TOV solution with the EoS of Fig.~\ref{F12} we can also link the radius of canonical stars to a soft EoS at densities
not much above the saturation point. This seems a more fundamental relation than linking the radii to the slope.
The cause and effect follows: the hyperon onset softs the EoS (at low densities), and this soft EoS produces the ``turn to the left''
in the mass-radius relation. The strong $\Lambda-\phi$ coupling produces a stiff EoS (at high densities), and this stiff EoS 
produces  very massive neutron stars.

\begin{figure}[ht] 
\begin{centering}
 \includegraphics[angle=270,
width=0.70\textwidth]{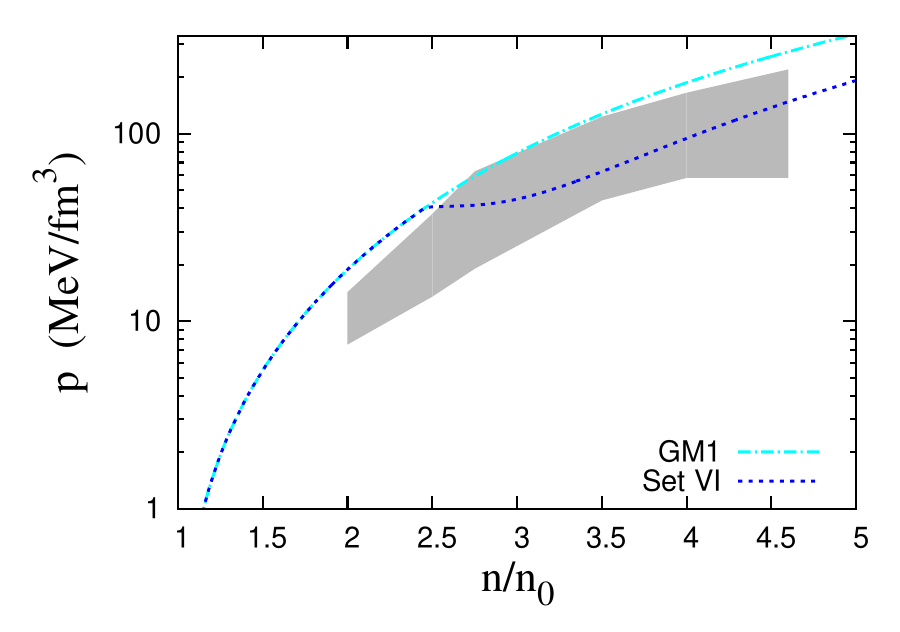}
\caption{(Color online) Role of the the $\Lambda$ threshold in the symmetric hypernuclear matter compared with
experimental constraint (hatched area). } \label{F10}
\end{centering}
\end{figure}

Another useful constraint to be assessed is the pressure of symmetric matter for densities up to five times the nuclear saturation density,
as inferred in ref.~\cite{Daniel} via HIC. As discussed in
ref.~\cite{Rafa2011,Lopes2013}, the GM1 is in disagreement with this constraint.
However, as pointed in ref.~\cite{Lopes2013} this could be due to the
fact we are not considering the onset of hyperons in the bulk of
nuclear matter. Now we define the symmetric hypernuclear matter as in
ref.~\cite{Lopes2013} and start by considering nuclear symmetric matter.
In this regime, by definition, the density of the protons is equal to the density of neutrons, $n_p = n_n$. Due to this,
the $\rho$ field is zero and the chemical potential of these particles are also the same, $\mu_n = \mu_p$. If we compress this
matter, the onset of strange particles, as $\Lambda$, becomes energetically favorable. Since ref.~\cite{Daniel} does not
rule out neither hyperons nor  even more exotic pictures, such as quark-hadron phase transitions, we assume the possibility
of $\Lambda$ onset in the symmetric matter, imposing:

\begin{equation}
\mu_p =\mu_n = \mu_\Lambda . \label{s5}
\end{equation}

This choice implies that only symmetric nuclear matter exists until the density is high enough so the creation of
strange particles becomes energetically favorable, softening the EoS.    The pressure of GM1 symmetric matter and symmetric hypernuclear 
matter alongside the inferred pressure up to five times nuclear saturation density (hatched area), obtained in ref.~\cite{Daniel} are plotted
  in Fig.~\ref{F10}. To not overcharge the figure, we again plot only sets D-IV and D-VI.

The hyperon threshold softens the EoS making experimental results from
HIC and theory reach an agreement again.  The hyperon threshold in symmetric hypernuclear matter happens
at 2.47 $n_0$ in opposition to 1.83 $n_0$ in beta stable matter. This is due to the absence of the $\rho$ field that increases the nucleon
chemical potential.
We see that the emergence of a new degree of freedom, besides being  able to explain 
the astrophysical observations of a very massive pulsar and the inferred low radius for 
the canonical mass, reconciles these results with those obtained in laboratory.
We conclude that our model with the onset of a new degree of freedom
agrees with astrophysical observations and HIC experiments.

\subsection{Threshold of an unknown particle}

The use of the $\Lambda$ hyperon as the new degree of freedom attaches several constraints:
 the mass and the $U_\Lambda$ are fixed, besides the $U_\Lambda^\Lambda(n_0/5)$
has a very small range. To extend our study of the onset of a new degree of freedom, in this section we use here, a yet not known particle,
that we just call M particle. We choose a mass lower than the $\Lambda$ hyperon to induce the onset of this new particle earlier than the obtained
with the $\Lambda$, as well  as the ``turn to the left". We check how
much we can reduce the radii of the canonical stars and yet produce massive ones.
We arbitrarily choose a mass of 1.040 GeV for this M particle, which
is an intermediate mass between the nucleon and the $\Lambda$ particle, a potential
depth $U_M$ equal to zero, meaning that this M particle does not
interact with the nucleon, and a value of -33 MeV  to the
  $U_M$ potential, what is similar to the $\Lambda$
hyperon potential. We also use two different approaches: a) moderate
values for the $M-\phi$ and $M-\sigma^*$  coupling constants, and  b) strong ones.  
The $M-\omega$ coupling is always zero to maximize the ``turn to the
left". The other values are present in Tabable~\ref{T9}.

\begin{table}[ht]
\begin{center}
\begin{tabular}{|c||c|c|c|c|c|c|}
\hline 
 Set &  $g_{M,\sigma}/g_{N,\sigma}$ &  $g_{M,\sigma^{*}}/g_{N,\sigma}$  & $g_{M,\phi}/g_{N,\omega}$ & $U_M$ (MeV) 
& $n/n_0$ ($Y_\Lambda = 10^{-4})$ & $n/n_0$ ($Y_\Lambda = 10^{-1})$ \\
  \hline
  E & 0.00& 2.86 & 2.39 & 0.0 & 1.62 & 1.91  \\
 \hline
  F & 0.00 & 2.86 & 2.34 & 0.0 & 1.62 & 1.86 \\
 \hline
  G & 0.117 & 2.83 & 2.39 & -33 & 1.26 & 1.50 \\
 \hline
H & 0.117  & 2.83 & 2.34 & -33 & 1.26 & 1.48  \\
 \hline
\end{tabular} 
\caption{Different sets varying the $U_M$ potential and the strength of the interaction.} 
\label{T9}
\end{center}
\end{table}

\begin{figure}[ht] 
\begin{centering}
 \includegraphics[angle=270,
width=0.70\textwidth]{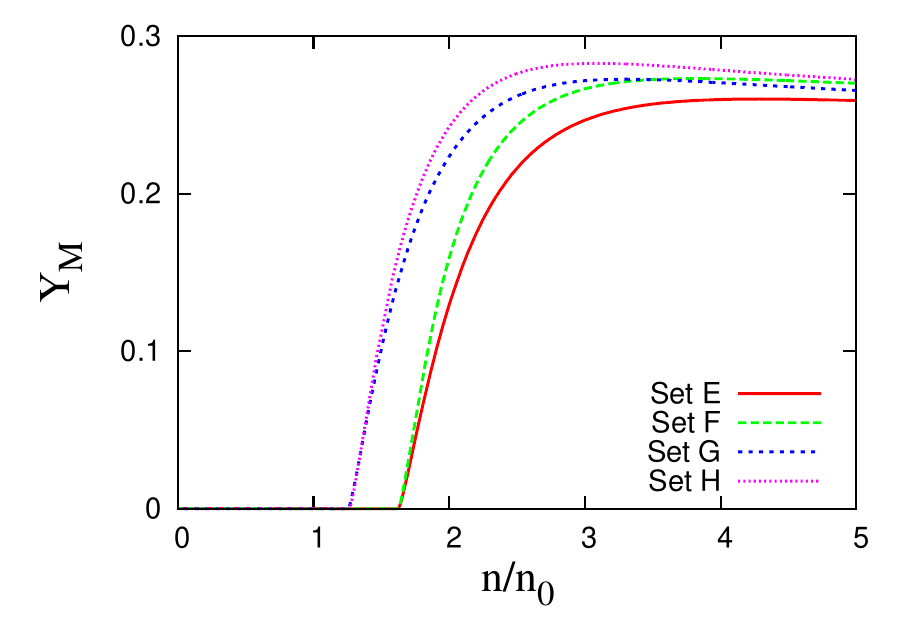}
\caption{(Color online) $\Lambda$ threshold and population for different values of the $U_\Lambda$ potential,$U_M$ potential and the strength of the interaction.} \label{F15}
\end{centering}
\end{figure}

 Due to the low mass of the M particle, the onset of this new degree of freedom happens at very low densities.
For a null potential, it happens at 1.62 $n_0$, while for a potential
of $U_M$ the threshold is at only 1.26 $n_0$, i.e., very close to the saturation point.
 
By analyzing  Fig.~\ref{F15}, we see that sets G and H, which  have the lower
potential depth,  causes the onset earlier than sets E and F, due to the contribution of two attractive channels, $\sigma$ and $\sigma^*$. The M fraction rises quickly, and $Y_M$ reaches 0.1 around 1.5 $n_0$. Therefore, in this model, not only the threshold happens at very low density, but also
the $Y_M$ becomes relevant at quite low densities. Due to this quick
grow of the $M$ particle population, 
the $\phi$ field dominates before two times the saturation density and
the $Y_M$ is supressed, never reaching 0.3. Within sets E and F, the
absence of the $\sigma$ field pushes the onset to hiher densities. But again, due to the strong $\sigma^{*}$ field, the $M$ population grows quickly and reaches 0.1 before two times the nuclear density. Again,  $\phi$ suppresses the $M$ particle at high
densities, never surpassing 0.3

\begin{figure}[ht] 
\begin{centering}
 \includegraphics[angle=270,
width=0.70\textwidth]{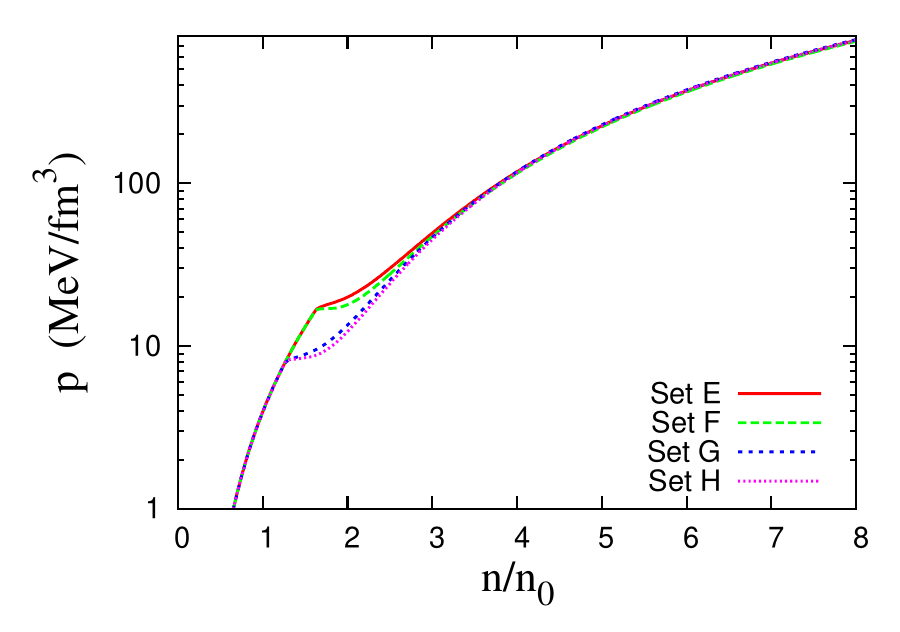}
\caption{(Color online) EoS for sets F and G.} \label{F16}
\end{centering}
\end{figure}

In Fig.~\ref{F16} we present the EoS for the four sets.
As expected from Table~\ref{T9},  set E is the stiffest  at low densities,
since it has only one attractive channel  and a strong $M-\phi$
coupling. Set H is the softest at low densities, due to the fact that in this set,
the M particle couples also to the $\sigma$ meson, which has a
non-zero value before the M onset.
Although very different at low densities, sets E and H have an almost indistinguishable EoS
at high densities. This explains similar neutron star maximum masses.

\begin{figure*}[ht]
\begin{tabular}{cc}
\includegraphics[width=6.0cm,height=7.0cm,angle=270]{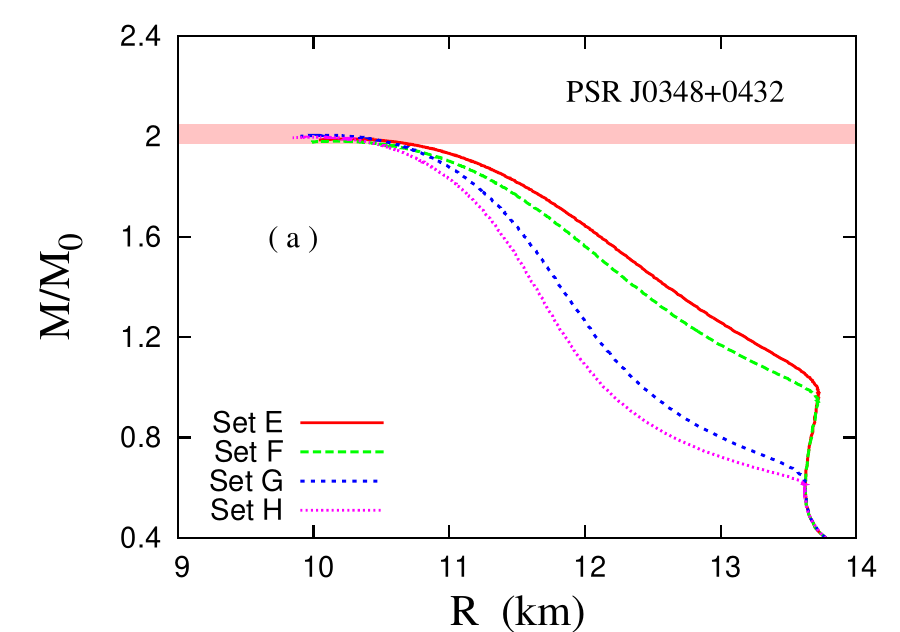} &
\includegraphics[width=6.0cm,height=7.0cm,angle=270]{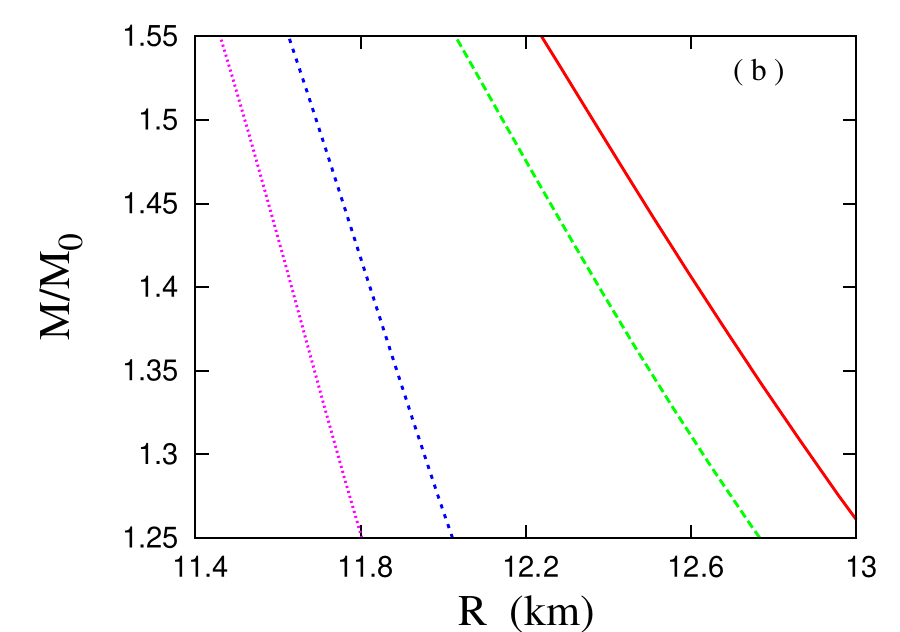} \\
\end{tabular}
\caption{(Color online) a) Mass-radius relation for different values
  of the $U_M$ potential, and different strength of interactions. b) Zoom in the mass region around $1.4M_\odot$.} \label{F17}
\end{figure*}

Now we plot the TOV solutions in Fig.~\ref{F17}.  As we already know,
the stronger the M potential depth, the earlier
is the onset of the new degree of freedom, softening the EoS. Consequently , the lower the star mass,
  the  earlier the ``turn to the left" happens.  For $U_M$ = -33 MeV the
  maximum neutron star mass before the onset of the M particle is only
  0.66$M_\odot$, while for a null potential depth, the maximum neutron star
mass without the M particle is 0.95$M_\odot$. Although significantly
different, we do not believe that such low mass neutron stars exist (a white dwarf is more 
probable). This indicates that if we believe that a yet unknown particle with a mass around one GeV is  present in the neutron star core, it would be in all known pulsars.

 The radius for the canonical 1.4$M_\odot$ varies from 12.59 km
  and 12.37 km for sets E and F to 11.80 km and 11.62 for sets G and H
  respectively. As sets G and H are softer at lower densities, these 
results corroborate our hypothesis that the star radii are linked to a soft EoS at low density.
Nevertheless, their radii are lower enough to be in agreement with the results presented in
ref.~\cite{Hebeler,Lim,Ozel,Steiner,Steiner2}, as well as their maximum masses with those described in ref.~\cite{Demo}.
All relevant properties are resumed in Table~\ref{T10}

\begin{table}[ht]
\begin{center}
\begin{tabular}{|c||c|c|c|c|c|}
\hline 
 Set &  $ M_{\max}/M_\odot$  &  $R_{M_{\max}}$ (km) & $n/n_0$ $(M_{max})$ & $R_{1.4M_\odot}$ (km) & $n/n_0$ $M(1.4M_\odot)$ \\
 \hline
 E & 1.99  & 10.28 & 7.83 &  12.59 & 3.47   \\
 \hline
  F & 1.98 & 10.18 & 7.96 & 12.37 & 3.61 \\
 \hline
  G & 2.01 & 10.09 & 7.93 & 11.80 & 3.73 \\
 \hline
  H & 2.00 & 10.04 & 7.95 & 11.62 & 3.82 \\
 \hline
\end{tabular} 
\caption{Main properties of neutron stars for different values of the $U_M$ potential and  strength of the interaction.} 
\label{T10}
\end{center}
\end{table}

\begin{figure}[ht] 
\begin{centering}
 \includegraphics[angle=270,
width=0.70\textwidth]{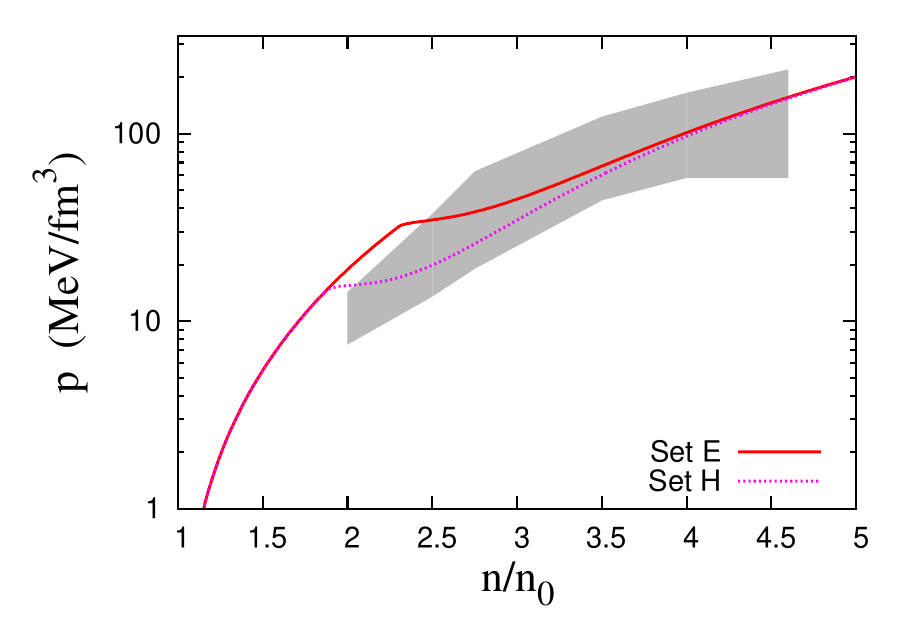}
\caption{(Color online) Symmetric hypernuclear matter compared with experimental
constraint (hatched area)  for different values of the $U_\Lambda$ potential and .} \label{F18}
\end{centering}
\end{figure}

To finish this last section we compare the results under the light of
the experimental determination of the pressure for symmetric matter as
pointed in ref.~\cite{Daniel}. In Fig.~\ref{F18} we present the two
more distinct sets E and H and we see that the with set E, as the M particle does not couple to $\sigma$, the EoS for the symmetric matter only enters in the experimental
region at 2.42 $n_0$. Nevertheless, the  $M-\sigma$ coupling in the H set, makes the EoS soft at low density and the EoS is in agreement 
for all experimental region, since it threshold happens at 1.96$n_0$. This fact indicates that the threshold of
the M particle, as the  new degree of freedom not only reduces the
star radii, but also reconciles experimental results and theoretical previsions.   

\section{Final Remarks \label{sec4}}

In this work we show how the onset of a new degree of freedom is able to reconcile
the recent measurements of very massive pulsars~\cite{Demo,Antoniadis}
with compact ones~\cite{Hebeler,Lim,Ozel,Steiner,Steiner2},
 and also satisfy HIC experimental constraints
 ~\cite{Daniel}. Choosing the $\Lambda$ hyperon as this new degree of
 freedom, we have shown 
 that we can satisfy the astrophysical and HIC experiments with the realistic values of the $U_\Lambda$ and $U_\Lambda^\Lambda(n_0/5)$ potentials.

We start by reviewing the effects of hyperon threshold in dense
nuclear matter and studying the influence of the coupling constant of the $\Lambda$ hyperon.
 The consequent softening of the EoS and  reduction of the maximum mass is a well-known theme in the literature~\cite{Glen,Glen2,Lopes2012}.
 The ``turn to the left" due to the hyperon threshold has also
   been noted in some works. For instance, in ref.~\cite{Bal99} 
the authors found that the hyperon  in the neutron star core could reduce the radius of the maximum mass neutron stars
 to values up to 3.4 km. However, they were not able to link the
 radius of the canonical mass to the hyperon threshold because in
 their work, the
 ``turn to the left" happens for masses above 1.5$M_\odot$. They could
 not explain the presently known 2$M_\odot$ pulsars either, once the maximum mass
 with hyperons they get are only 1.8$M_\odot$. In the same way
 ref.~\cite{SB2002} found that very attractive hyperon-hyperon
interaction could produce radii as low as 8 km. Yet, again the maximum
mass found is below 1.8$M_\odot$.
Similar results are also presented in ref.~\cite{Novak,Ortel}. Using the so called LS220 model with only one 
new degree of freedom (as we did) the authors found  maximum masses of 1.91$M_\odot$ and 1.95$M_\odot$ if this new degree
is a $\Lambda$ hyperon or a pion respectively. In both cases the radii of the canonical stars are below 12.5 km.
However LS220  is a non-relativistic model~\cite{Ortel} and the masses are slightly below the experimental limit for the
PSR J0348+0432.

 We show that for the well-established $\Lambda$ potential depth of -28 MeV, 
in the traditional $\sigma\omega\rho$ model, there is little influence of the hyperon
 onset on the radii of stars around the canonical mass of
 1.4$M_\odot$, besides the fact that most parametrizations are unable
 to explain massive pulsars as the PSR J0348+0432. To overcome this
 issue we have added a new repulsive channel, through
the strange vector $\phi$ meson. This allows us to stiffen the
EoS and increase the maximum mass without changing any properties of
nuclear matter and with little effect on the hyperon threshold and
also to construct a stiff  EoS, that is able to explain the massive
PSR J0348+0432. However, although there is a 
small reduction of the radius of the canonical star, it is not enough
to be in agreement with those discussed in
 ref.~\cite{Hebeler,Lim,Ozel,Steiner,Steiner2},

In the third part of our work we add a new attractive channel, the
strange scalar $\sigma^{*}$ meson  as in~\cite{Miy88,Fortin95}.
 This new attractive field softens the EoS at low densities, making the ``turn to be left" more
pronounced. This  compresses  the subsequent stars, allowing us to
construct an EoS that predicts  maximum masses around 2.0$M_\odot$ , while the canonical 1.4$M_\odot$ has a radius as lower as 12.45 km.
 In other words, the $\sigma^{*}$ meson,
allows us to construct simultaneously massive and compact stars. 
 We are also able to link the low radius to a soft EoS at low
 densities, what seems a more fundamental reason than relating it to
 the symmetry energy slope. Furthermore, we are able not only 
to predict correctly the $U_\Lambda$ = -28 MeV, but the
$U_\Lambda^\Lambda(n_0/5)$ lies in the range between describe in the ref.~\cite{Fortin95} and ref.~\cite{Ahn2013}.
Moreover, the emergence of a new degree of freedom again reconciles
this EoS with HIC experiments.

In the last section, we explore a more general result, evoking an yet
not known dark matter particle with mass between
the nucleon and the $\Lambda$ hyperon. We explore two different
potentials and different interaction strengths and we are able to
still predict massive and very compact neutron stars, with the canonical mass of only 11.62 km.

Before summarizing our findings, it is important to stress that
  specific density dependent models (TW \cite{tw},  DDH$\delta$
  \cite{gaitanos} and
  DD-ME$\delta$ \cite{roca-maza} can produce massive stars alongside 
  canonical stars with radii lower than 12.6 Km if only neutrons and
  protons are included in the EOS  and even
  smaller radii, in between 11-12 Km, if hyperons are considered
  \cite{mariana2}.  Nevertheless, in this case, the maximum mass drops
  approximately 10\%. 

Another important point worth mentioning refers to the relation
between the equation of state used to describe the star crust and its
radius. A comprehensive study was performed in  \cite{cp2016} and it
is clearly seen (and also stated by the authors) that the uncertainty
on the radius is connected with the properties of the EoS. This means
that a $\omega-\rho$ interaction, which can be used to control the
values of the symmetry energy and its slope \cite{Rafa2011} can also be
used to further improve the results presented here. Moreover, the
inclusion of the pasta phase in the inner crust also modifies the
radii of the canonical stars, as shown in \cite{guilherme, lena}.
Despite it, it is worth to mention that there is still some 
uncertainty about the neutron stars radii as pointed in ref.~\cite{Steiner3, Miller}

 The use of a new degree of freedom to compact neutron stars with
  canonical mass produces an additional 
constraint: the radii of the 2$M_\odot$ in all models are now around
10 km. New observations of such massive pulsars
are required in order to validate our proposal.

We finish this work by pointing out some empirical facts either
  developed or reinforced through out it:
a) The EoS must be soft at low densities to predict compact neutron stars; b) A new degree of freedom can
 soften the EoS at low density and still produce stiffen EoS at high densities, explaining the massive PSR J0348+0432 and the measurements
of low radii pulsars; c) This new degree of freedom can be the $\Lambda$ particle with its realistic $U_\Lambda$ and $U_\Lambda^\Lambda(n_0/5)$;
d) The new degree of freedom  not only
fulfills b),  but also reconciles the results with those obtained 
with HIC experiments; e) All the models used are thermodynamically consistent without violate the Le Chatelier principle.
f) The use of the new degree of freedom to compact the neutron stars also constraint the radii of 2$M_\odot$ pulsars.

\vspace{.5cm}

{\bf Acknowledgments;} DPM acknowledges partial support from 
Conselho Nacional de Desenvolvimento Cient\'ifico e Tecnol\'ogico
(CNPq) and Project INCT- FNA Proc. No. 464898/2014-5 and
LLL from CEFET/MG. LLL also informally dedicates this work to the 
memory of Professor Stephen W. Hawking (1942-2018), whose book - {\it O universo numa casca de noz}
(The Universe in a Nutshell) -  inspired the author, at the young age of 15, to embrace the path of theoretical physics.

\end{document}